\documentclass {rspublic}
\usepackage[dvips]{graphicx,color}
\usepackage{cite,subfigure,amsmath}
\usepackage{harvard}

\usepackage{amssymb, amsthm, alltt, setspace,bbm}

\definecolor{Red}{rgb}{1,0,0}
\definecolor{Blue}{rgb}{0,0,1}

\begin{document}
\title[Lattice-gas simulations of dynamical geometry]{Lattice-gas simulations of dynamical geometry in one dimension}
\author[P. J. Love,  B. M Boghosian, D. A. Meyer  ]{ Peter J. Love$^1$,  Bruce M. Boghosian$^1$, David A. Meyer$^2$}
\affiliation{$1$ Department of Mathematics, Tufts University, Medford, Massachusetts 02155 USA, $2$ Department of Mathematics, University of California/San Diego, La Jolla, CA 92093-0112, USA }
\maketitle

\begin{abstract}{Dynamical geometry, lattice-gas, nonequilibrium dynamics}
We present numerical results obtained using a lattice-gas model with dynamical geometry defined by~\cite{bib:meyer7}. The (irreversible) macroscopic behaviour of the geometry (size) of the lattice is discussed in terms of a simple scaling theory and obtained numerically. The emergence of irreversible behaviour from the reversible microscopic lattice-gas rules is discussed in terms of the constraint that the macroscopic evolution be {\it reproducible}. The average size of the lattice exhibits power law growth with exponent $1/2$ at late times.  The deviation of the macroscopic behaviour from reproducibility for particular initial conditions (``rogue states'') is investigated as a function of system size. The number of such ``rogue states'' is observed to decrease with increasing system size. Two mean-field analyses of the macroscopic behaviour are also presented.

\end{abstract}

\section{Introduction}

Lattice-gas automata (LGA) have successfully modelled a variety of systems including multiphase and multicomponent flow, chemical self-assembly and flow in porous media~\cite{bib:adler,bib:rk,bib:lcb2,bib:bce}. Transformed into discrete quantum systems, suitable generalisations of LGA have successfully modelled the Dirac and Schr\"odinger equations~\cite{bib:bogwash,bib:meyer2}.  All of these applications possess a {\sl fixed\/} background geometry which is represented in the LGA models by a fixed lattice (and boundary conditions).  In contrast, there are many natural systems with {\sl dynamical\/} geometry.  

Only one lattice-gas model with dynamical geometry has so far been studied. Hasslacher and Meyer defined the only reversible lattice-gas automata with dynamical geometry in one dimension~\cite{bib:meyer7}. We are interested in modelling geometry {\sl intrinsically}, rather than with reference to some ambient space in which the system is embedded. Since it is the lattices in LGA which define the geometry, the model of Hasslacher and Meyer allows collisions in which lattice sites may be created and deleted. The restriction that the lattice-gas rules be reversible in time\footnote{By which we will always mean {\em local\/} reversibility.} restricts us to a single model in one dimension provided we consider only two lattice vectors at each site. The classical, local, deterministic, single species LGA with two vectors per site and static geometry in one dimension is completely trivial as it can be interpreted as consisting of particles which simply move to the left or right without change even upon scattering. Nevertheless, the one dimensional LGA with dynamical geometry is significantly more interesting~\cite{bib:meyer7} . 

As it is the constraint of time reversibility which uniquely specifies the collision rules for the model we shall consider, it is worth considering carefully what time reversibility means in the context of the LGA. The evolution rule in all LGA has two phases.  First each particle {\sl advects\/} to the lattice point along its lattice vector. Second, the particles at each lattice point {\sl collide\/} according to some local rule.  The advection phase is trivially reversible and so it only remains to make the collision step reversible. In the deterministic models considered here this is achieved by making the collision a permutation amongst the states of the lattice site with the same values of the conserved quantities (in our case, mass and momentum). Physical time reversibility is not achieved by inverting all particle momenta and then running the forward evolution rule. Although advection and scattering are both reversible, these operations do not commute. Exact time reversibility is achieved by applying the inverse of propagation and collision in reverse order. 

\begin{figure}[htp]
\centering
\includegraphics[width=0.8\textwidth]{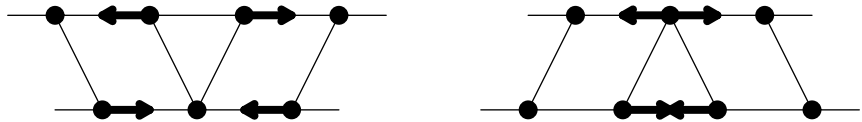}
\caption{ The scattering rules for our
dynamical geometry LGA. Time runs up the page, and the lattice before and after the collision is depicted. Solid black dots indicate the lattice sites. The presence of a particle and its direction are indicated by  an arrow. An edge is created when two particles advect to
the same lattice point and an edge is deleted if after advection each 
endpoint is occupied by a single particle with outward momentum. Spectator particles are allowed in the empty vectors depicted here provided their inclusion does not violate the exclusion principle.}
\label{fig:latrules}
\end{figure}

In the one dimensional dynamical geometry model the same advect/collide formalism for particle evolution is retained, but the collision rules allow for local changes of the lattice.  An edge is created when a vertex splits into two vertices, and so the most local rule would depend only on the particle configuration at a single vertex.  The exclusion principle restricts these configurations to consist of no particles, one particle, or two particles with opposite momenta. As explained in~\cite{bib:meyer7} the constraint of reversibility allows only the possibility of splitting a vertex which is occupied by two particles, as shown in Figure~\ref{fig:latrules}. Reversibility then implies the dual edge deletion rule also shown in Figure~\ref{fig:latrules}.  These rules are well defined as they cannot have overlapping domains of application.  They define the simplest lattice-gas model with dynamical geometry.

In keeping with the theme of this issue, we consider this model from the perspective of microscopic {\em vs}. macroscopic dynamics. When one seeks closed, effective macroscopic descriptions for systems out of equilibrium several problems arise. It is possible to define many macroscopic variables in terms of the plethora of microscopic degrees of freedom. It is often unclear which variables are necessary or sufficient to obtain a closed macroscopic description. We expect such macroscopic descriptions to be irreversible in general, by contradistinction with fundamental reversible microscopic laws. Finally, the domain of validity of macroscopic descriptions and when one may expect significant deviation from the ``average'' behaviour is often unclear. The model considered in this paper provides a simple example with which to consider some of these issues. 

The remainder of the present paper is organised as follows. In Section~\ref{sec:macbe} we consider in very general terms the dynamics of the size of the lattice and introduce the idea of reproducibility of the non-equilibrium dynamics of the geometry to motivate the numerical studies of Section~\ref{sec:sim}. We then analyse the model using two mean-field theory approaches in Section~\ref{sec:mft} before closing the paper with some conclusions and future directions.

\section{Macroscopic behaviour}\label{sec:macbe}

Having defined our reversible lattice-gas model, we now wish to consider what its macroscopic behaviour will be. The ``geometry'' of the lattice in one dimension is trivial, the only macroscopic variable characterising the geometry is the size of the lattice. The time evolution of the size of the lattice will be dependent on the (reversible) microscopic evolution of the particle content. Denoting the size of the lattice and the microscopic state at time $t$ by $L(t)$ and $s(t)$ respectively we may write the two evolution equations as:
\begin{equation}
\Sigma_i(0) \ni s_i(0)  \mapsto s_f(t) \in \Sigma_i(t)
\end{equation}
\begin{equation}
L_i(0) \mapsto L_f(t), 
\end{equation}
where $\Sigma_i(0)$ is the set of microstates corresponding to initial size $L_i$ and $\Sigma_i(t)$ is the set of microstates resulting from $t$ actions of the collision rule on each member of $\Sigma_i(0)$. Because the dynamics is deterministic the number of elements in $\Sigma_i(0)$ and $\Sigma_i(t)$ is the same. We now define the property of exact {\it reproducibility} of the evolution of $L(t)$. The set of microstates consistent with the new lattice size $L_f(t)$ is $\Sigma_f(t)$, and exact reproducibility is the statement that {\it all} initial conditions with size $L_i(0)$ evolve to a final state with size $L_f(t)$. In terms of the elements of the phase space this statement is: 
\begin{equation}\label{eq:subset}
\Sigma_i(t) \subseteq \Sigma_f(t)
\end{equation}
 For a lattice of size $L$ containing $p$ particles there are ${2L \choose p}$ initial conditions, and hence~(\ref{eq:subset}) implies:
\begin{equation}
{2L(0) \choose p} \leq {2L(t) \choose p},
\end{equation}
implying that if the evolution of the lattice size is exactly reproducible, then the evolution is irreversible:
\begin{equation}
L(0) \leq L(t). 
\end{equation}

Clearly exact reproducibility is a very strong condition, and it is straightforward to construct microstates which violate this condition. We can, however, make a weaker statement based on the behaviour of a distribution over the initial states. If we consider an initial distribution which is sharply peaked around $L_i(0)$, the weaker statement (which we will refer to as {\em reproducibility}) is that at later times the distribution will be sharply peaked around $L_f(t)$. The argument above is repeated exactly, but now becomes a statement about the average lattice size over many independent realisations, and we must include the possibility that there will be ``rogue states'' in the tails of the distribution which lead to macroscopic behaviour far from the average. However, as the system size increases we expect the fraction of such microstates to decrease, until in the ``thermodynamic limit'' the probability of serious deviations from the macroscopic evolution is vanishingly small. This argument is presented and discussed in detail in~\cite{bib:garrett}.

Having assumed that a simple macroscopic evolution rule for $L(t)$ exists, and shown that its existence implies such a rule must be irreversible, we may now try and obtain the form of such a rule. In our numerical studies of the model we may therefore verify three things. Firstly, does the model geometry in fact possess a well defined reproducible behaviour? Secondly, if the mean behaviour is well defined, what form does it take? Thirdly, we expect there to be ``rogue'' microstates which will deviate from the macroscopic evolution. We can therefore explicitly count the number of such states as a function of increasing system size.
\section{Simulations}\label{sec:sim}

\begin{figure}[htp]
\centering
{\includegraphics*[width=0.8\textwidth]{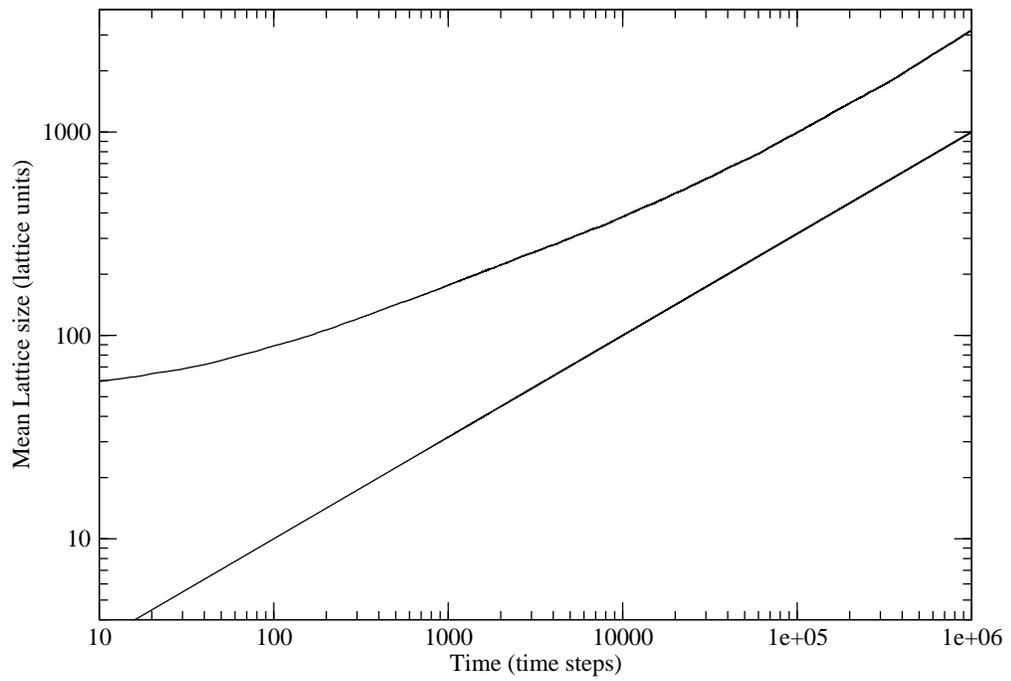}}
\caption{Upper line: Ensemble average lattice size $L(t)$. The lower line is a power law with exponent $0.5$ and is included as a guide to the eye only.}
\label{fig:mean}
\end{figure}

Simulations were performed in order to obtain the average time evolution of the lattice size $L(t)$. The system was initialised with $25$ particles on a periodic lattice with $50$ sites. The initial state of the system was chosen randomly, and $1000$ different initial conditions were simulated for $10^6$ time steps. The memory requirements for each realisation were sufficiently modest that individual simulations could be performed on a single processor. The large number of realisations required in order to obtain good statistics meant that the simulations were distributed across a $16$-node Athlon Beowulf cluster. The lattice size was stored at each time step for each separate simulation. The mean lattice size as a function of time is shown in Figure~\ref{fig:mean}.
The mean lattice size as a function of time was fitted to a power law function given by Equation~(\ref{eq:fit1}).
\begin{equation}\label{eq:fit1}
L(t)=at^b
\end{equation}
The first $150000$ timesteps were not included in the fit as this early time behaviour was not well described by a power law. The fits were obtained using the Levenberg-Marquardt algorithm and the errors on fit parameters were obtained by a careful {\em ex post facto} analysis of the behaviour of $\chi^2$ near the minimum. The exponent obtained was $0.505\pm0.02$, for a fit with $\chi^2=0.47$, consistent with a power law growth with exponent $1/2$.

\begin{figure}[htp]
\centering
\includegraphics[width=0.8\textwidth]{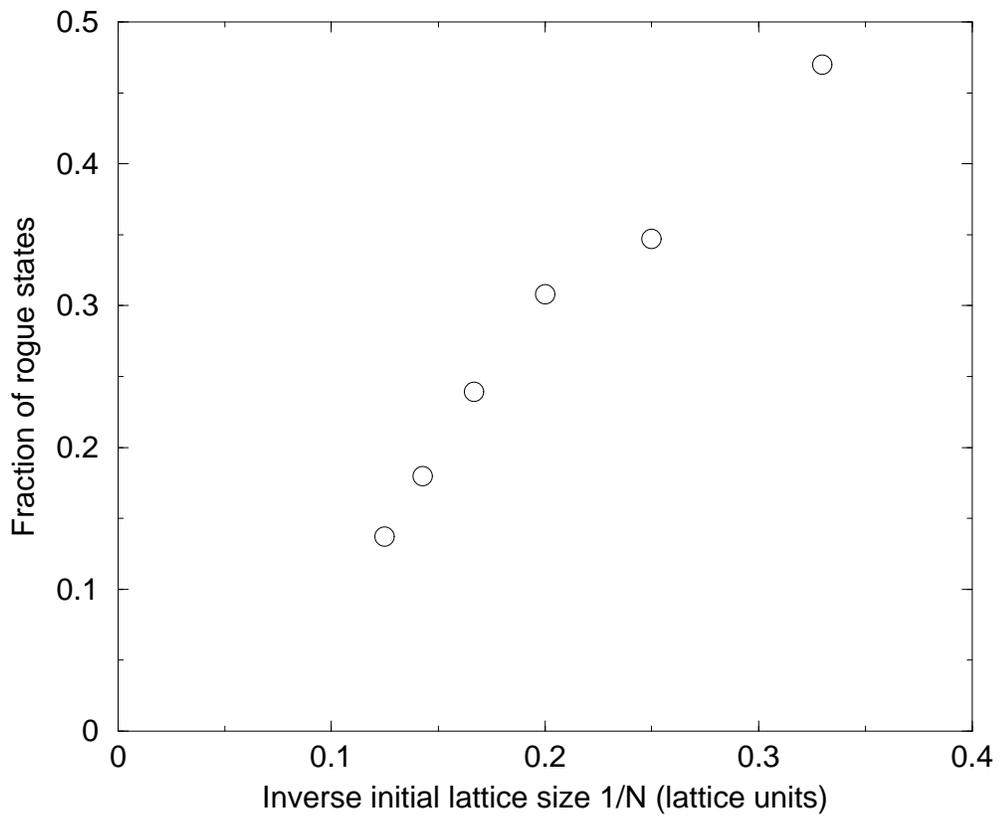}
\caption{Fraction of rogue microstates as a function of inverse initial lattice size for lattice sizes (from right to left) $3,4,5,6,7,8$.}
\label{fig:rogue}
\end{figure}

The final numerical investigation we can perform is to search for the presence of ``rogue'' microstates and study their behaviour with increasing system size/ number of particles. In order to test the conjecture that the number of such rogue states will decrease with increasing lattice size, the time evolution of all microstates for initial lattice sizes $3,4,5,6,7$, and $8$ was computed for $10,000$ time steps. Two types of rogue states were found, states in which the lattice size remained constant, and states in which the lattice size oscillated. The fraction of rogue states (defined as those systems with lattice size less than $100$ at time step $10000$) was computed in each case and the results are plotted in Figure~\ref{fig:rogue}. These results are consistent with the number of rogue states becoming vanishingly small in the limit of large lattice size.

\section{Analysis}\label{sec:mft}
\subsection{Mean-field theory - particles}

We consider one particular lattice site $x$. The probability that this site splits into two at the next timestep is:
\begin{equation}
Pr(x \textrm{ splits at next timestep })=Pr((x-1, +) \textrm{ occupied and } (x+1,-) \textrm{ occupied} )
\end{equation}
where $(x,\alpha)$ denotes a lattice site and momentum pair. If the density of right/left moving particles is $\rho_{\pm}$, the ``mean-field'' approximation is:
\begin{equation}
Pr(x \textrm{ splits at next timestep })=\rho_+\rho_-
\end{equation}
then the expected number of splits is $E[+] = L\rho_+\rho_-$, where $L$ is the number of lattice sites. Similarly, consider one particular edge $e$, between lattice sites $x$ and $x+1$. Then
\begin{equation}
\begin{split}
&Pr(e \textrm{ deleted at next timestep })=Pr((x, +) \textrm{ occupied and } (x+1,-) \textrm{ occupied} \textrm{ and } (x-1, +)\\ 
&\phantom{Pr(e \textrm{ deleted at next timestep })=}\textrm{ not occupied and } (x+2,-) \textrm{ not occupied})
\end{split}
\end{equation}
Making the same mean-field approximation as above we obtain:
\begin{equation}
Pr(e \textrm{ deleted at next timestep })=\rho_+\rho_-(1-\rho_+)(1-\rho_-)
\end{equation}
Combining the expected number of splits and contractions gives:
\begin{equation}
\begin{split}
E[\Delta L]&= E[+]-E[-]\\
&=  L\rho_+\rho_-(\rho_++\rho_--\rho_+\rho_-)\\
\end{split}
\end{equation}
 writing $\rho_{\pm}=n_{\pm}/L$ gives:
\begin{equation}
\begin{split}
E[\Delta L]&=\frac{p}{L^3}\left(Ln-p\right)\\
\end{split}
\end{equation}
Where $p=n_+n_-$ and $n=n_++n_-$. We can therefore write a difference equation for $E[L_t]$:
\begin{equation}
\begin{split}
E[L_{t+1}]&=E[L_t]+E[\Delta L]\\
&=E[L_t] + \frac{p}{E[L_t]^3}\left(E[L_t]n-p\right)\\
\end{split}
\end{equation}
We can convert this to a differential equation:
\begin{equation}
\frac{dL}{dt} = \frac{p(nL-p)}{L^3}
\end{equation}
and integrate to obtain:
\begin{equation}
\begin{split}
\left[\frac{(L-p/n)^3}{3} + \frac{3(L-p/n)^2n}{2p} +\frac{3(L-p/n)n^2}{p^2}\right]
+\frac{n^3}{p^3}\ln\left( L-p/n\right)&= pnt +C\\
\end{split}
\end{equation}
This approach predicts an asymptotic growth of $t^{1/3}$. This disagrees with our numerical data which shows an asymptotic growth of $t^{1/2}$. The mean-field theory approach taken here is the natural thing to do if one is familiar with other types of lattice-gas model, where one defines average quantities which undergo macroscopic evolution. However, the lattice-gas model underlying our dynamical geometry model is non-interacting, and does not possess a local equilibrium. The average density for the model without dynamical geometry does not posses a macroscopic evolution which can be obtained via the Chapman Enskog expansion. The validity of utilizing such mean quantities in the dynamical geometry case is therefore open to question.

Secondly, the replacement of two-particle correlations with products of one-particle densities is appropriate if the collision frequency is rather high so that a particle rapidly becomes uncorrelated with neighbouring particles. In the dynamical geometry case, however, the density is a decreasing function of time and so this approximation, which underlies the mean-field theory, gets progressively worse at later times. 

\subsection{Mean-field theory - lattice}

We have seen that a mean-field theory treatment based upon the average particle densities fails to reproduce the simulation results. As one goes further into the asymptotic regime the picture of an average density of particles becomes inappropriate because eventually the lattice is so large that the particles are very well separated. A treatment of the model in terms of the dynamics of the geometry, {\em i. e.} in terms of the particle separations, becomes appealing.

Particle separations have two properties which determine their time evolution. They can be increasing, decreasing or constant, and they can be odd or even. From the point of view of the dynamics of the total lattice size, only decreasing separations are of interest. We see immediately that there is a natural timescale for the dynamics of separations. Immediately after a collision has occurred, the time to the next collision is given by half the size of the smallest decreasing separation. At the next collision the lattice will increase or decrease by one site. We introduce an ensemble with $N$ members in which we label each realisation by $i$. If we denote the time to the next collision of the $i$th realisation by $\Delta T =L_i/c_i$ where $L_i$ is the size of realization $i$, and $c_i>1$, then:
\begin{equation}
L_i(T+\Delta T) = L_i(T)+\left(N^{-e}_i-N^{-o}_i\right)
\end{equation}
where $N^{-o}_i$ ($N^{-e}_i$) is the number of odd (even) decreasing separations of size $L_i$ in realisation $i$. Dividing by $\Delta T = L_i/c$, averaging over all members of the ensemble:
\begin{equation}\label{eq:dldt}
E\left[\frac{L(T+\Delta T)-L(T)}{\Delta T}\right] =E\left[\frac{c_i(N^{-e}_i-N^{-o}_i)}{L_i}\right]
\end{equation}
From our discussion of the connection between macroscopic reproducibility and irreversibility in Section~\ref{sec:macbe} we know that if the behaviour of the lattice size has a reproducible evolution, a necessary condition for the expectation value on the left hand side to have any meaning at all, the lattice size will be increasing in time on average. We therefore know that if the expectation value on the left hand side of equation~\ref{eq:dldt} is a meaningful quantity then the right hand side will be positive.  
We now sum over $i$ and introduce the number of increasing,decreasing and odd/even links in the ensemble via the numbers $N^{AB}$, where $A\in\{+,0,-\}$ and $B\in\{o,e\}$. For $N$ members of the ensemble and with $P$ particles we have:
\begin{equation}
NP = \sum_{A\in\{+,0,-\}}\sum_{B\in\{o,e\}} N^{AB}
\end{equation}
Using the above and replacing averages of products by products of averages we have 
\begin{equation}
E\left[\frac{L(T+\Delta T)-L(T)}{\Delta T}\right]= P\frac{E[c_i]}{E[L]} -\frac{ \left(2N^{-e}+N^{+o}+N^{+e} +N^{0e} +N^{0o}\right)}{N}\frac{E[c_i]}{E[L]}
\end{equation}
As every quantity in the second term is positive, and the left hand side must be positive, the first term must be dominant here. We therefore neglect the second term and convert the above to a differential equation:
\begin{equation}
\frac{dL}{dt} = E[c_i]P\frac{1}{L}, 
\end{equation}
giving an asymptotic scaling of one half, consistent with the numerical data.
\section{Conclusions}

We have performed extensive simulations in order to characterise the late-time macroscopic behaviour of a one-dimensional lattice-gas model with dynamical geometry. We have adduced two mean-field arguments to explain the asymptotic growth. A mean-field theory based on the Boltzmann approximation for the particle densities yields an incorrect prediction of $t^{1/3}$ growth. A similar argument applied to the separations of particles yields the correct exponent. 

As this model was originally motivated by the desire to describe interfaces in a two-dimensional multiphase fluid intrinsically, the value of the growth exponent for the lattice size deserves some discussion. In binary immiscible fluids undergoing phase separation into droplet phases, three growth exponents have been observed. An exponent of $1/3$ is characteristic of a diffusive Lifshitz-Slyzov growth mechanism, and exponents of $1/2$ and $2/3$ have been observed in simulations including hydrodynamics. It is worth pointing out that two-dimensional lattice-gas simulations of droplet growth display a growth exponent of $1/2$, where more mean-field oriented lattice-Boltzmann simulations obtained growth exponents of $1/3$~\cite{bib:weig,bib:wagneryeo,bib:gonoryeo}. This is at least an amusing coincidence, but we mention this mainly to caution the reader against reading too deeply into it. The physical interpretation of the lattice-gas particles in the one-dimensional  model when considered as degrees of freedom of an embedded interface is unclear at best.

This model possesses a generalisation to two dimensions which we are actively pursuing. As may be readily imagined, the two-dimensional model is considerably more complex. This is due in no small part to the presence of non-trivial local curvature. Work in progress includes characterising the hydrodynamic behaviour in a fixed, curved background geometry and producing efficient code for the simulation of dynamical geometry in two dimensions.     

\section*{Acknowledgements}

PJL would like to thank the American University of Beirut, Tufts University Department of Mathematics and the EPSRC for funding support enabling him to attend the 12th International Conference on Discrete Simulation of Fluid Dynamics 2003. PJL would also like to thank the local staff and students of the American University of Beirut for their excellent hospitality. Simulations were performed on the Tufts University Department of Mathematics 16-node Athlon cluster obtained under AFOSR grant number [F49620-01-1-0456]. PJL was supported by the DARPA QuIST program under AFOSR grant number [F49620-01-1-0566]. BMB was supported in part by the U.S. Air Force Office of Scientific Research under grant number [F49620-01-1-0385]. DAM was supported in part under AFOSR grant F49620-01-1-0494. The authors would like to thank Jeff Rabin for productive discussions.


\end{document}